\documentstyle[12pt]{article}
\begin{document}
\baselineskip=24pt
\setcounter{page}{1}
\parskip=0pt plus2pt
\textheight=22cm
\begin{titlepage}

\begin{center}
    {\Large \bf 
{Transport coefficients at Metastable Densities from models of
Generalized Hydrodynamics }}
\end{center}

\vspace{0.2in}

\begin{center}
{\it
Sudha Srivastava and Shankar P. Das
\\
School of Physical Sciences, Jawaharlal Nehru University
\\New Delhi 110067, India.}

\end{center}
\vspace*{1.0in}

\begin{center}
{\bf ABSTRACT}
\end{center}
\noindent
In the present work we compute the enhancement in the long time
transport coefficients due to correlated motion of fluid particles at
high density. The fully wave vecor dependent extended mode coupling
model is studied with the inclusion of an additional slow variable of
the defect density for the amorphous system. 
 We use the extremely slow relaxation of the density
 correlation function observed in the light scattering experiments
 on colloids to estimate the input parameters for the model
The ratio of 
long time to short time diffusion coefficient is studied around the
the peak of the structure factor. 
\noindent

\vspace{1in}
\noindent
PACS number(s) : 64.70P, 05.60, 64.60C
\end{titlepage}
\newpage

\section{Introduction}
 Liquids undercooled below
 the freezing temperature without crystallization shows a
 sharp increase in their transport properties like viscosity
 and acquire solid like properties {\it e.g.} has a finite
 shear modulus.
 The dynamics of the dense liquids are studied 
 in terms of generalized transport coefficients,
 expressed in the form of memory functions which account
 for the correlated motion of the fluid particles at
 high density.
  Frequency and wave vector dependence of these quantities,
  relevant for behavior over different time and
  length scales are studied through models of the liquid 
 obtained from the statistical mechanics of a many particle system.
 Much progress has been made in this field through a self-consistent
  approximation of the memory function in terms of the slowly
 decaying modes that exist as a consequence of microscopic
 conservation laws .
 This has been termed in the
 literature as the mode coupling theory (MCT)
 \cite{kimmaj,sid-trans}.
A key result of the simple mode coupling model
 \cite{leshouc,mcth} for the memory function,
 is a two step relaxation process involving the power law decay
 of correlation over intermediate time scales crossing over to the
 stretched exponential behavior in the long time, termed as the
$\alpha$-relaxation.
 With increase of density, the liquid become
very viscous and relaxation slows down strongly.
The basic driving mechanism for the enhancement of the
 viscosity comes from a feed back to the transport coefficient
 from the coupling of the slowly decaying density correlation
 functions.
In the simple model, long time correlation of density fluctuations
 freeze completely beyond a critical density and the
fluid undergoes a  dynamic transition to a nonergodic phase.
An extended version \cite{D+M,gtz87,jimrudi}
of the mode coupling theory with proper coupling of the density
 and current modes demonstrated that there is a final
exponential relaxation of the density correlation
function over asymptotic time scales.
The final decay process which shifts to longer time scales
 with density can be interpreted as a process similar
to the diffusion of free volumes in the amorphous system whereby
ergodicity is maintained.

In the theoretical formulation of the dynamics of the supercooled 
 liquids often extra slow modes \cite{MPP,cohen1,cohen2} in addition to 
 the usual ones that follow from simple  hydrodynamics, has been considered.
Extension of the simple MCT has been
 done \cite{phlet,plaw} along these lines to include the extra
 slow mode of defect density that develop in the amorphous solid
like state.
In Ref. \cite{dsch} and \cite{bkim} the displacement field $\vec{u}$
 was used to describe the extra hydrodynamic mode due to the solid like
nature of the system. Ref. \cite{bkim} demonstrated that due to the
nonlinear couplings of the density fluctuations and the displacement
field $\vec{u}$ in a solid like state, the renormalized expression
 for transport coefficients include terms linear in density correlation 
 function.
 Such terms in the memory function appears
 over time scales in which
 the defect correlation is taken to be a constant.
 Due to this phenomenological inclusion of the
 extra slow mode of the defect density, a line of transition
  is obtained instead of a single critical density.
 In this formulation of the extended MCT,
temperature or density  dependent exponents of power law relaxation \cite{plaw}
 are obtained.
 Over the longest time scale the defect correlation
 as well as the density correlation function do however
  decay to zero and thus restores the ergodic behavior.
 In a recent work \cite{phlet} we have studied the coupled
 dynamics over longest time scale,
 through a schematic model dropping all wave vector dependencies.
 The asymptotic behavior for both these correlation functions
 have been considered there to study the relaxation over various time
regimes.
 However, in order to study the transport
 behavior with respect to change in the thermodynamic parameters
 like density, the wave vector 
 dependence of the  model has to be taken into account. 
 In the present work we consider 
 the wave vector dependent model for the dense supercooled liquid
 and  the coupling of the
 density fluctuations to the extra slow mode of defect density
 are included.
 The dynamics of the density correlation function
 is considered up to the time scales where the final
 ergodicity restoring mechanism present.
 It was demonstrated \cite{D+M} from a
 non perturbative approach that the self consistent feed
 back mechanism allows a non-zero value of the cutoff function.
 However, explicit evaluation of the strength of the function is
 extremely difficult beyond the one loop order.
 On the other hand in certain simple systems like colloids
 very slow relaxation of the structure function has indeed
 been observed. This indicate that the cutoff function
 is effective over very long time scales.
 In the present calculation we model this through a self-consistent
 picture where the defect density and the density fluctuations
 over longest time scales behave in a similar way.
 We use here this criteria to
 close the dynamical equation for the density correlation
 function - instead of actually evaluating cutoff function
 self consistently.
 The cutoff function is estimated to give decay of the structure
 factor that agrees with experimental results on hard sphere like 
 systems like Colloids.
 We then compute the mode coupling integral to obtain the
 renormalized transport coefficient as function of density.
 The coupling of the slowly decaying density
 fluctuations to the defect density in the super cooled state
 is a key aspect of this model and allows a physical picture
 for the ergodicity restoring mechanism.

 The plan of the paper is as follows. In the section II we give
 a brief description of the model studied and the approximations
 involved while in section III we describe our results for
 the transport coefficients. We end the paper with a small
 discussion.

\vspace*{1cm}

\section{Transport Coefficients at High Density}
\subsection{Longitudinal Viscosity}
In this section we describe briefly the model used in computing
 the longitudinal viscosity or the sound attenuation constant.
The key quantity of interest is the  dynamic structure factor or the
density auto correlation function (DACF).
The Fourier-Laplace transform of the
DACF $\psi_q(t)$ normalized with respect to its equal time value \cite{D+M},
 defined as,
\begin{equation}
\label{lap-t}
\psi_q(z) = -i \int_0^{\infty} e^{izt} \psi_q (t) dt
~~~~~~~~~~~~~~~ Im(z) > 0
\end{equation}

\noindent
In the basic formulation of the MCT, this quantity is expressed in terms of
 the generalized memory function ${\Gamma}_q(z)$,
\begin{equation}
\label{dacf}
\psi_q(z) = {{z+i {\Gamma}_q(z)}
\over
{ z^2 - \Omega_q^2 + i {\Gamma}_q (z)[z+i\gamma_q(z)]}} .
\end{equation}
\noindent
$\Omega_q$ corresponds to a characteristic
microscopic frequency for the liquid state dynamics.
The memory function which is the generalized longitudinal
viscosity, consists of the bare and mode coupling part,
${\Gamma}_q(z) =  \Gamma^0_q + \Gamma^{mc}_q(z)$.
$\Gamma^0_q$ is related to bare or short time dynamics
with uncorrelated collisions. The mode coupling contribution
$\Gamma_q^{mc}(z)$ signify the  correlated motion in the dense liquid
 and can be expressed in terms of the coupling of density fluctuations
 dominant at supercooled densities.
 Several models has been used to compute the Generalized memory function
 for the supercooled state. The simplest model \cite{mcth,beng}.
 express the memory function in terms of a bilinear product of the 
 density correlation functions i.e. 
 $\Gamma^{mc} (t)$ $ = $ $c_2 {\psi}^2(t)$ .
 The model that contains a linear term of the density correlation function
 in addition to the quadratic one, 
 i.e.  $\Gamma^{mc} (t)$ $ = $ $c_1 \psi (t) + c_2 {\psi}^2(t)$,
 has been referred to in the literature 
 \cite{leshouc,alba} $\phi_{12}$ model.
 In the wave vector independent form this model gave rise to the stretched 
 exponential relaxation. 
 In later works it was demonstrated that by considering the solid
 like nature of the amorphous state at high density 
 such a linear term can be obtained in the
 renormalization of the transport coefficients. The set of hydrodynamic
 variables were now extended to include the slowly relaxing
 defect density. The resulting non linearities in
 the equations of Generalized Hydrodynamics gives rise to the above
 memory function in terms the DACF.
 In order to consider the very dense system approaching the glass transition 
 we will use here this model for the dynamical behavior.
 Following Ref. \cite {yeo-maj,yeo}, the mode coupling part of the 
 generalized transport coefficient  is given by,
\begin{equation}
\label{mem}
{\Gamma}_q^{mc}(t) = \int \frac{d{\vec k}}{n_0 (2\pi)^3}
S(q) [\tilde{V}^{(1)}(q,k)\psi_k(t) \phi_{k_1}(t) + 
\tilde{V}^{(2)}(q,k)\psi_k(t) \psi_{k_1}(t)]
\end{equation}
\noindent
with $\vec{k_1} = \vec{q} - \vec{k} $. Here $\phi_{k_1}(t)$ is the
 normalized defect density auto correlation function and gives the
 contribution to the transport coefficient coming from the
 coupling of the defect density with the density fluctuations.
The mode coupling  vertices with full wave
vector dependence 
are  given by
\begin{equation}
\label{v1}
\tilde{V}^{(1)}(q,k)=  2y  \tilde{U}(q,k)S(k)S(k_1)
+ \kappa n_0c(k)S(k) 
\end{equation}
\noindent
and
\begin{equation}
\label{v2}
\tilde{V}^{(2)}(q,k)= \frac{1}{2} \tilde{U}^2(q,k)
- y \tilde{U}(q,k) S(k)S(k_1)
\end{equation}
\noindent
The vertex function $\tilde{U}$ in (\ref{v2}) is given by,
\begin{equation}
\label{u}
\tilde{U}(q,k) = \frac{n_0}{q} \big[ \hat{q}.k c(k) +
\hat{q}.(k_1)c(k_1) \big]
\end{equation}
\noindent
where $n_0$ is the particle number density. 
It is assumed  that the defects
are moving in a metastable double well potential 
 and this well depth is characterized by $y$.
The defect density is assumed here to be
a variable similar to mass density and is
weakly interacting with the latter.
$\kappa$ is a dimensionless parameter characterizing the coupling of
the defect density with the particle density in the Free energy functional
corresponding to the stationary state around which fluctuations are considered.
For $\kappa$ and $y$ both equal to zero, (\ref{mem}) reduces to the
simple MCT result, \cite{beng}.
If the defect correlation is assumed to be very long lived
 compared to the time scales over which initial power
 law relaxation persists, $\phi_{k_1}(t)$ in equation (\ref{mem}) can be taken
 as a constant and the so called $\phi_{12}$ model results on 
 ignoring all wave vector dependencies. 
 The density dependence of the power law
 exponents and the resulting dynamic instability in the model has been
 studied in Ref. \cite{plaw}. Here we will consider the time scale
 over which both density correlation and the defect correlation
 decays. 
The quantity $\gamma_q(z)$ in the R.H.S of eqn. (\ref{dacf})
 has crucial implications for the asymptotic dynamics \cite{D+M}.
 If $\gamma$ is ignored then a sharp transition in the supercooled
 liquid to an ideal glassy phase results  beyond a critical density. 
The Laplace transform of the density correlation function develops 
 a $1/z$ pole.  
 This is a widely studied \cite{sid-trans} model for the dynamics of 
 supercooled liquids.
 However, with the presence of $\gamma_q$ at high density when 
 $\Gamma$ gets large, the pole shifts to $1/(z+\gamma_q )$.
 It has been demonstrated
  \cite{D+M,jimrudi} that in the small $q$ and $\omega$ limit, the
 quantity $\gamma_q$ $ \sim$ $ q^2 \gamma$, where $\gamma$ remains finite, 
 instead of self consistently going to zero. 
 This implies a diffusive decay of the
  density correlation restoring ergodicity in the longtime limit.
The quantity $\gamma$ that provides a mechanism that cuts off the
sharp transition of the fluid to an ideal glassy phase
is $O(k_BT)$ \cite{D+M} to leading order in the perturbation theory.
It is a consequence of the coupling of the
density and current correlation in the compressible fluid.
Formal expression for $\gamma$ was obtained in self consistent form in 
ref. \cite{D+M} in terms of current correlation function.
 However beyond the one loop order explicit evaluation of the strength of 
 the cutoff function becomes extremely complicated. 
 In the present work, we estimate this by interpreting
 it as similar to slow diffusion of free volumes in the
 dense liquid.
 We use the simple model of defect correlation as 
  $\phi_q(t) = e^{-\delta (q) t} $ and replace
  the $\gamma(q) $ in equation (\ref{dacf}) by $\delta (q) $ to have a
 closed equation for the density correlation function.
The density correlation function is obtained by solving 
equation (\ref{dacf}) in the time space \cite{spd-utrecht}.
The quantity $\delta$
is treated as a parameter and is adjusted to obtain agreement in the
 asymptotic behavior of the dynamic structure factor as seen in experiments.
 For this purpose the data from light scattering on colloids by Pusey {\it et al.}
 is  \cite{van-pusey} is considered. 
 The solution of the MCT equations provide the result for the 
 relaxation of the density correlation function with time scaled
 w.r.t  the microscopic time scale for the Hard Sphere system - namely
 the Enskog time.
 The present analogy of the Hard Sphere system with the colloids
 is used only with regard to the nature of the cooperativity.
 The unit of time 
 $\tau$ in terms of which the experimental results are
 presented, then needs to be related to the characteristics 
 microscopic time scale of the Hard Sphere model i.e. $\sqrt{\beta m} \sigma$. 
$m$ is the mass of the liquid particle and $\beta$ is the Boltzmann factor. 
The latter is related to the Enskog time \cite{yip} for the system.
 The MCT models for the dynamics we consider here does not describe
the short time dynamics or the bare transport coefficient for Colloids -
 the relevant comparison is over the long time scales or the nature
of the correlated motion of individual units. These correlated motions 
are quantified in terms of the mode coupling terms.
 Thus the comparison of the theoretical data with the results
 of the experiment  is done upto an undetermined scaling
 factor of the short time scales in the two systems.
 The decay of density correlation as a function of time
 is evaluated over a suitable wave number grid.
 The generalized transport coefficients expressed in terms
 of the coupling of density correlations from the mode coupling 
 contribution is then evaluated. 
The longitudinal viscosity is computed relative to its bare or short 
 time contribution.
The dependence of the Generalized Viscosity can be probed in the
 present model from various approach - namely the time dependence
 as well as the wave vector dependence of the long time limit.
 The detailed results from the numerical solutions are presented
 in the next section. 

\vspace*{.5cm}

\subsection{Long time Diffusion}
 The density fluctuation in a dense fluid at finite wave number 
 close to the peak of the static structure factor
 follows a diffusive behavior.
 Once again similar to the viscosity the
 diffusion constant has short
 time value $D_s$ and the long time $D_L$ behaviors. 
 If all kinds of correlated motions
 are completely ignored then the diffusion coefficient is related
 to the bare transport coefficient. At high densities the cooperative
 nature of the dynamics is essential in computing the diffusion
 coefficient. This is obtained by including in the memory function
 the mode coupling contribution.
 The short time dynamics depends on the microscopic nature of the
 system. We use this diffusion constant to define the unit of time
 namely $\tau = \sigma^2 /D_s$
 The mode coupling part is computed through the self-consistent
expression in terms of the density correlation function.
 For the DACF, the model described 
 in the first part of this section is used. 
 Once again instead of focussing on the microscopic model we
 compute the ratio of the longtime to short time diffusion coefficient.
This ratio of diffusion coefficients for the density fluctuations 
at finite wave number is expressed in terms of the memory function 
 through the formula
\begin{equation}
\label{difu-mem}
\frac{D_L(q)}{D_s} = \frac{1}{s(q) \tilde \Gamma_q}
\end{equation}
\noindent          
where $\tilde\Gamma_q=\int_0^{\infty} d\tau \Gamma_q^{mc}(\tau)$,
is the long time limit of the memory function given
by equation (\ref{mem}) and s(q) is the static structure factor.
The colloidal time ($\tau$) is related to the Enskog time 
$t_E$ through a constant $\Delta$, {\it i.e.}
 $\tau=\Delta t_E$. 
The numerical results for the transport coefficients obtained
 from these formulations w.r.t to density and wave number
 variations are described in the next section.

\vspace*{1cm}
\section{Results for the transport coefficients}
The dynamic density-density correlation function $\psi_q(t)$ is obtained as a 
solution of the nonlinear equation 
obtained by inverse transforming equation (\ref{dacf}). 
The generalized transport coefficient $\Gamma_q(t)$
has two parts namely the bare and the mode coupling part.
The bare quantities in the Hard Sphere system for  which the
dynamical equations are solved are obtained in terms of standard
 models of kinetic theory and the unit of time is expressed in
 terms of the Enskog time \cite{yip}.
The integro-differential equation obtained from equation (\ref{dacf})
is solved for $\psi_q(t)$ self-consistently over a wave vector grid of 
$N=200$ and an upper cutoff of $\Lambda\sigma=30$ using numerical 
integration. Here we have used the 
Percus-Yevick \cite{PY} solution with Verlet-Wiess \cite{VW} correction 
for the hard sphere static structure factor.
The constant $\Delta$ relating the two time scales, colloidal and the 
microscopic time, is estimated from the dynamic structure factor obtained
from colloidal experiments.
The colloidal time ($\tau$) is estimated as $2.9 \times 10^5$ times Enskog 
time ($t_E$).  
As discussed in the previous section the 
cutoff function $\gamma$ in equation (\ref{dacf}) 
is estimated through a parameter $\delta$
which is inversely
related to the relaxation time for the defects. The parameter $\delta$ is
adjusted to get an agreement with the light scattering data, reported in Fig. 5(a) of Ref.
\cite{van-pusey} for the dynamic structure factor, in the asymptotic
limit. 
The light scattering data supplied in Ref. \cite{van-pusey} was thus used to
 obtain the corresponding value of the parameter $\delta$ at that density.
In Fig. 1 the $\delta$ values thus obtained in units of $\sigma^2/\tau_H$
are plotted as a function of
packing fraction for the dense fluid. 
Here $\tau_H=10t_E$ is the hard sphere
unit of time with respect to which the time is scaled. 
For intermediate densities, the
 $\delta$ value  follows a power law behavior approaching divergence 
 at $\eta\approx 0.536$ as shown by dotted line. 
 With increase of density the viscosity smoothes off to a slower
 enhancement. 
The parameter $\delta$ as estimated from the asymptotic behavior of the
dynamic stucture factor of the colloids,
is then used to compute the density correlation function
 from the model equations self consistently.
 This is also used to compute the memory function over very
 long time scales and the transport coefficients in the supercooled
 state is thereby computed in terms of the Mode Coupling contributions
 accounting for the correlated motions at  higher densities.
The long wavelength limit of the memory function $ \Gamma_q(t)$, 
reduces to the form,
\begin{equation}
\label{visd}
\lim_{q{\rightarrow 0}}\Gamma_q(t) = \frac {1}{\beta m} 
\int \frac{d\vec{k}}{{(2\pi)}^2} 
\{ \tilde{V}^{(1)}(k)\psi_k(t) \phi_{k_1}(t) +
 \tilde{V}^{(2)}(k)\psi_{k}(t) \psi_{k_1}(t) \} 
\end{equation}
\noindent
with the vertex functions given by,
\begin{equation}
\label{v10}
\tilde{V}^{(1)}(q{\rightarrow 0},k)\equiv 
\tilde{V}^{(1)}(k) =
 2c(k) \{ 2yS(k)+\kappa \} S(k) + \frac{2}{3}c_1(k) S^2(k)
\end{equation}
\noindent and in a similar way,
\begin{equation}
\label{v20}
\tilde{V}^{(2)}(k)=
 \big[  c^2(k) + \frac{1}{5}{c_1}^2(k) +\frac{2}{3}c(k)c_1(k)
- 2yc(k) - \frac{2y}{3}c_1(k)\big] n_0 S^2(k) \ \ \ .
\end{equation} 
\noindent
where $c_1(k) = k c^{\prime}(k)$.
In Fig. 2 the generalized memory function $\Gamma(t)$ in
the long wavelength limit is plotted as a function of 
time for the packing fraction $\eta=0.542$.
 Here the time $(\tau_H)$ is expressed in the units of the Enskog time,
$t_E$. The corresponding value of $\delta$ is $1.0 \times 10^{-7}$ in units
of $(\sigma^2/\tau_H)$ and 
the metastability parameters in the MCT equation are 
$\kappa=-0.01, y=0.12$ respectively. 
The metastability parameter $y$ determines the potential well used for
 the defect density. In figure 3 the form of the well used
  in the calculations for figure 2 is shown.
 We have expressed the result for $\Gamma(t)$
in a dimensionless quantity in terms of the unit ${(\beta m)}^{-1}$.
$\Gamma (t)$ also follows the two step relaxation process.
At intermediate time scales long compared to the microscopic time
 scales, it follows a power law decay  finally crossing over to the 
 stretched exponential decay at longer times. 
The variation of the stretching exponent $\beta$ as a function of packing 
fraction is shown in Fig. 4. With 
increase in density $\beta$ decreases implying increase or
 stretching of relaxation of the DACF at higher densities. 
This critical slowing down of the system  is estimated through the zero
 frequency limit of the longitudinal viscosity. 
 In Fig. 5 longitudinal viscosity scaled with 
respect to the corresponding bare transport coefficient,
 $\Gamma (z=0)$ is plotted as a function of packing fraction. 
Over the low density region the viscosity shows a 
power law behavior approaching a divergence around $\eta=0.538$. 
The corresponding exponent is $1.1$.
With increase of density an abrupt increase in the 
viscosity by approximately $4$ orders of magnitude is observed. 
Thus at the onset of the glassy regime, a relatively strong enhancement 
in viscosity is obtained from the self-consistent mode coupling models.  
We also present the wave number dependence of the
 viscosity to illustrate the nature of transport over different length
 scales. 
Indeed an analysis of the self consistent expression
 for the density correlation function indicates that the relevant quantity
 for long time decay is inverse of $S(q)\Gamma_q$. This is plotted  in
Fig. 6 scaled with respect to the bare time $v_o \sigma$ , where $v_o$ 
is the thermal velocity  of the particles. Here it is shown as a 
function of the wave number $q\sigma$ for $\eta=0.542$. 
The values for
the metastability parameters $\kappa$ and $y$ are same as stated for 
figure 2. 
The result shows that the time scale indicating
decay rates for different wave numbers is slowest near the peak of the
 structure factor and for small wave number it approaches a constant
 value. It should be noted that this is the final decay process in the
 case when one completely ignores the cutoff mechanism. Thus with increase of
 density this will go into a delta function type peak. However the
 existence of the cutoff mechanism ensures the proper ergodic behavior
 through the diffusive process over longest time scales \cite{fischer}
 At very small wavelength i.e. large wave numbers the transport coefficients
are same as calculated from the short time properties of the liquid,
{\it i.e.}, $\Gamma_{q} \rightarrow 1$.
This indicate that over small length scales the 
effect of correlated motion is insignificant even at very high density
and the mode coupling has negligible implications on the transport properties. 
With increasing wavelength the 
viscosity increases and is maximum at $\lambda \rightarrow \infty $ or 
$q\sigma \rightarrow 0$. The signature of the structure of the 
liquid are seen in terms of very small maximas and minimas (see Fig.
6) over the whole wave vector range.
Fig 7 shows the ratio of long time to short time diffusion coefficient 
at the peak of the structure factor as a function of density of
the supercooled system. It shows a
power law decay of diffusion at intermediate densities which shows a 
divergence around $\eta=0.538$ with an exponent close to unity.
As the density increases further the
decrease in the diffusion coefficient slows down due to the presence of
cutoff effect in the present model. The diffusion coeffients are calculated in
colloidal units with $\Delta = 2.9 \times 10^5$.
To study the dynamics at the microscopic time scales we have calculated the
ratio of the long time to short time diffusion coefficient as a function of
wave vector.
Fig 8, presents the behavior of the Diffusion Coefficient over
a wave vector range around the diffraction peak. The results shown
correspond to the packing fraction $\eta$=.542. We observe that the diffusion
coeffient is minimum at the peak of the structure factor where maximum
number of nearest neighbours are present. We observe qualitatively similar
behavior as observed by Pusey {\it et al.} \cite{segre}. The values of the metastability
parameter used are same as mentioned above for figure 2.
It should be noted
 that it is only over the finite wave vector range one can describe
 the density fluctuations following a diffusive mechanism even in the
 models where there is no cutoff mechanism of hydrodynamic diffusion
 is considered. Of course in this case it will only be vaild upto densities
 below the ideal transition point beyond which transport coefficient
 at all wave numbers would diverge in the ideal nonergodic state. 
 However, with cutoff mechanism taken into
 account in the present work, as a motion of defect density  - one
 is not restricted to densities below the ideal glass instability
 only. Fig. 8  considers the effect on the diffusive process
 at finite length scale within the self consistent formulation of
 the problem where the cutoff mechanism is also being taken into 
 account.

\section{Discussion}
The standard mode coupling theory of glass transition predicts a 
sharp dynamic transition  with diverging transport coefficients. 
However in realistic system such a freezing does not occur through 
substantial slowing down of the ralaxation as seen in the light
scattering data of colloids.

  In the present work we considered a phenomenological model for
   the dynamics of the supercooled system where structure factor
   for the hard sphere system was used as an input for thermodynamic
   properties. The density correlation function is the key quantity
   in the mode coupling model. 
In this formulation the self-consistent form of the cutoff function
used is obtained  in terms of the coupling to current correlation 
functions to lowest order in perturbation theory.  
The present model includes the phenomenological description of slowly
decaying defects in the amorphous structures. We interpret the cutoff
mechanism of final decay of density correlation function as being
related to the diffusion of free volume or defect in the amorphous
system. The strength of the cutoff function is estimated by seeking an
agreement with the relaxation of dynamic structure factor seen in
experimental studies on colloidal systems.
The correlation of the density fluctuations in the presence of very
slowly decaying defect correlations contribute to the transport
coefficients calculated using the feed back mechanism.
The present theory of course does not compute the bare transport
coefficient which is related to the short time scale dynamical
behavior. The mode coupling contributios are computed relative to the
short time or the bare properties. The longitudinal viscosity in the
long time limit is computed and shows an enhancement by four orders of 
magnitude over small density range with an approach to power law 
divergence around a critical range of density $\rho=1.028$ 
which finally smooths off to a slower decay.
The wave vector dependence of the transport coefficient is also
considered.  The decay rate over finite wave vector shows a minimum at
the peak of the structure factor.

The relaxation of density fluctuations over the intermediate length
scales follows a diffusive behavior \cite{biman,jim}. Due to
effects of correlated motions in the fluid particles at high density
the diffusion coefficient is changed over long times. We compute this
by taking into account the Mode Coupling effects due to slow decay of
density correlations. Thus the ratio of the long time to short time
diffusion constant is calculated showing that the diffusion process slows
down. However the diffusion process is fater than the final
diffusive decay of defect density fluctuations given by $\delta$ in the $q
\rightarrow 0$ limit. The latter is related to the ergodicity restoring
mechanism in the fluid and correspond to the largest time scale in the
dynamics. 

The present work demonstrates that the very slow decay of DACF
seen in colloids can be understood in terms of the extended MCT and the
corresponding values of the transport coefficients like longitudinal
viscosity or long time diffusion constant is calculated in terms of the
short time properties of the system.
\vspace*{1cm}

\section*{Acknowledgements}
SPD acknowledges support from the NSF grant INT9615212.
SS acknowledges the financial support from the UGC, India.

\vspace*{1cm}

\newpage
\section*{Figure Captions}
\noindent

\subsection*{Figure 1}
Plot of $\delta$ in units $\sigma^2/\tau_H $ Vs packing fraction.
\noindent

\subsection*{Figure 2}
The normalized generalized viscosity  $\Gamma(t)$  in units of
$(\beta m)^{-1}$ as a
function of time at $q\sigma=0.0$, $\eta=0.542$, $y=0.15$, $\kappa=-0.01$ and 
$\delta=1.0 \times 10^{-7}$.
\noindent

\subsection*{Figure 3}
Variation of depth of the potential with $n/n^*$ for
$\eta_c$
 0.542. 
In the figure $h^*(n)$
represent the dimensionless quantity $h(n)\beta
\epsilon n \sigma^3$. 
The metastability parameters used are $\kappa=-0.01$ and $y=0.12$.

\subsection*{Figure 4}
Variation of $\alpha$-relaxation stretching  exponent  $\beta$ with 
packing fraction.
\noindent

\subsection*{Figure 5}
Plot of the longitudinal viscosity $\Gamma$ scaled w.r.t to the
bare transport coefficient as a function of packing fraction $\eta$.
\noindent

\subsection*{Figure 6}
variation of $\Gamma_q^{-1}$ in units of $v_0 \sigma$ with wave number 
$q\sigma$ at $\eta=0.542$.
\noindent

\subsection*{Figure 7}
Variation the longtime to short time diffusion constant $(D_L/D_s)$ as a
function of packing fraction.

\subsection*{Figure 8}
Wave vector dependence of the ratio of longtime to short time diffusion
constant, \\ $D_L(q^*)/D_s \times 10^5$ near the peak of the structure factor.
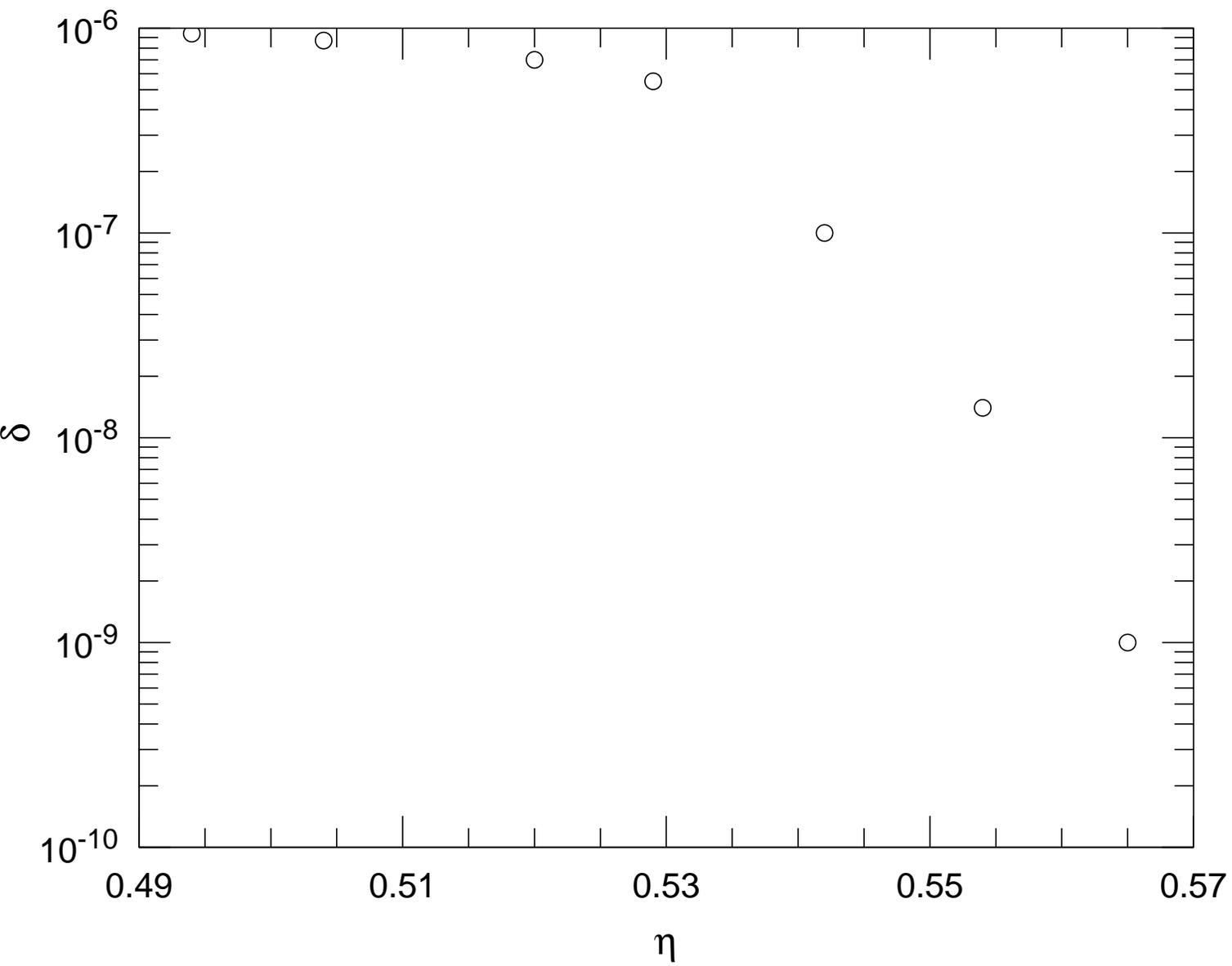
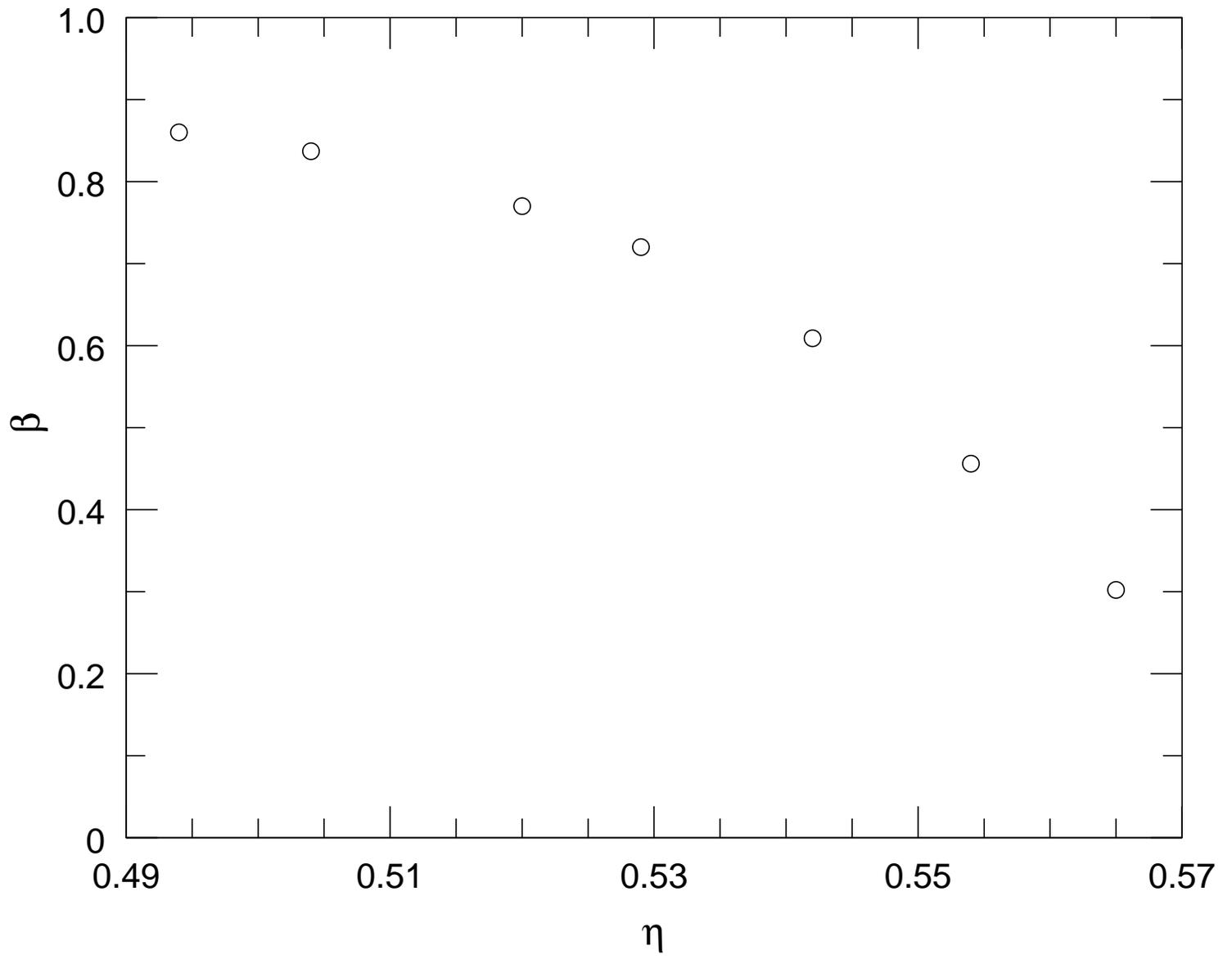
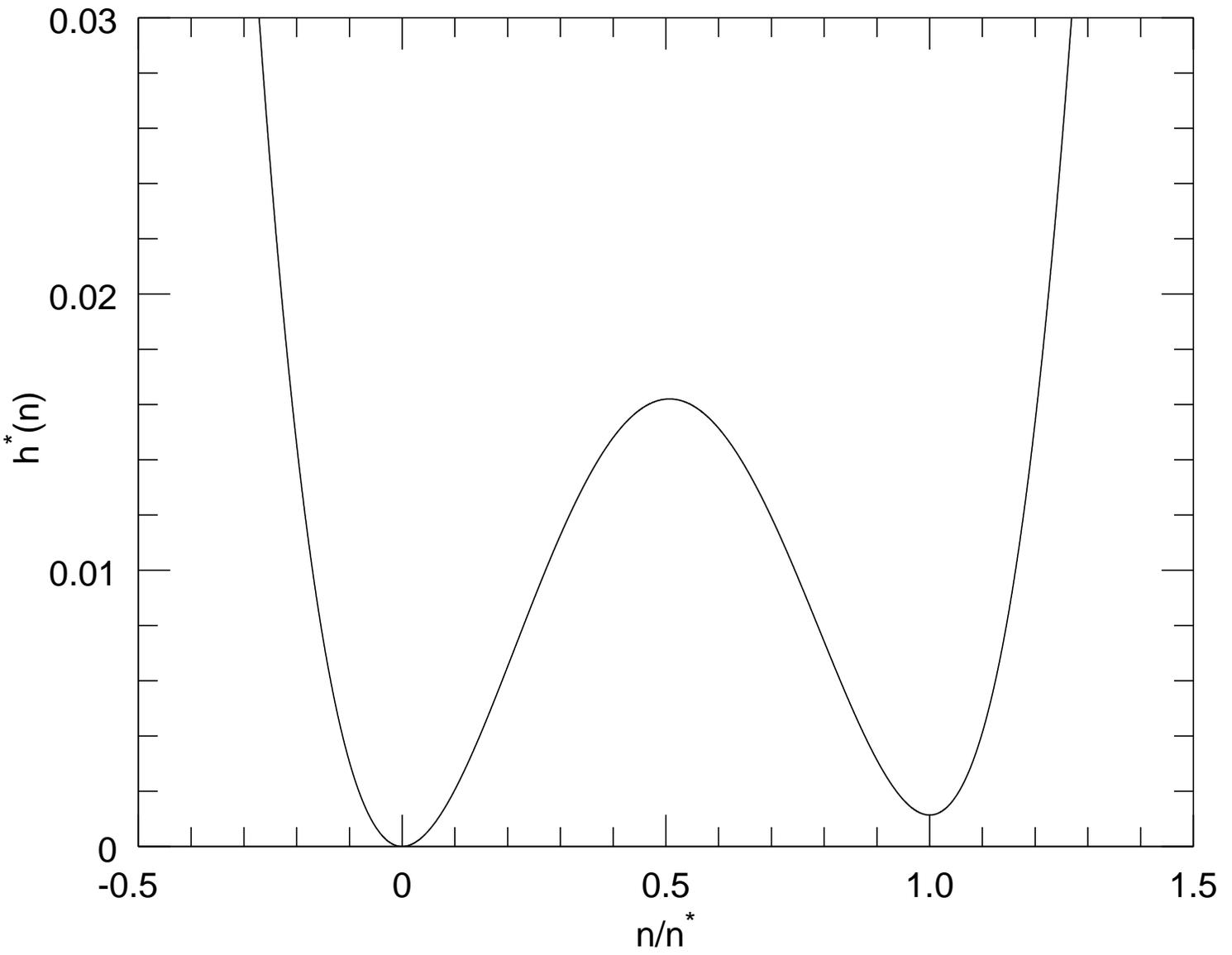
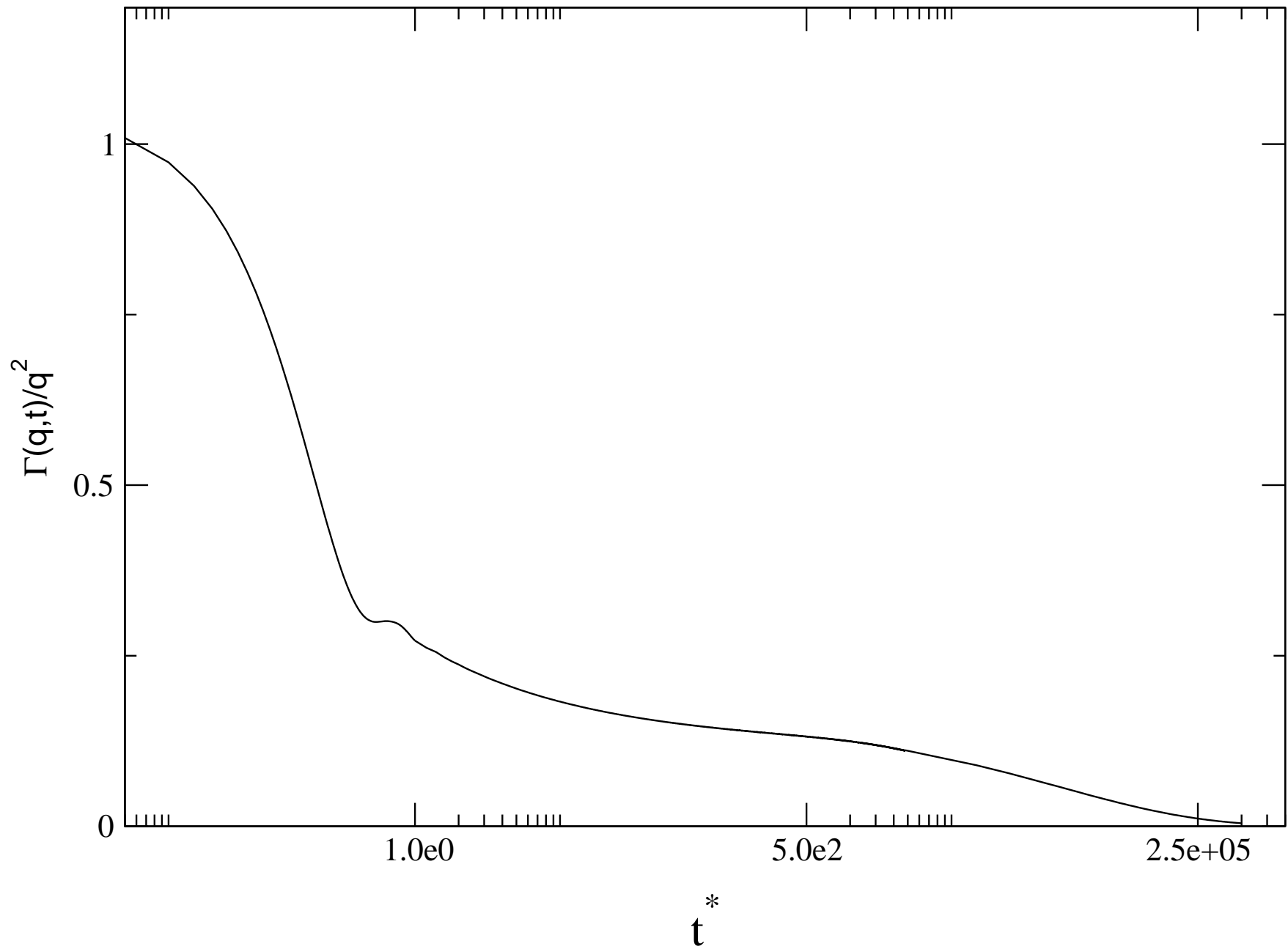
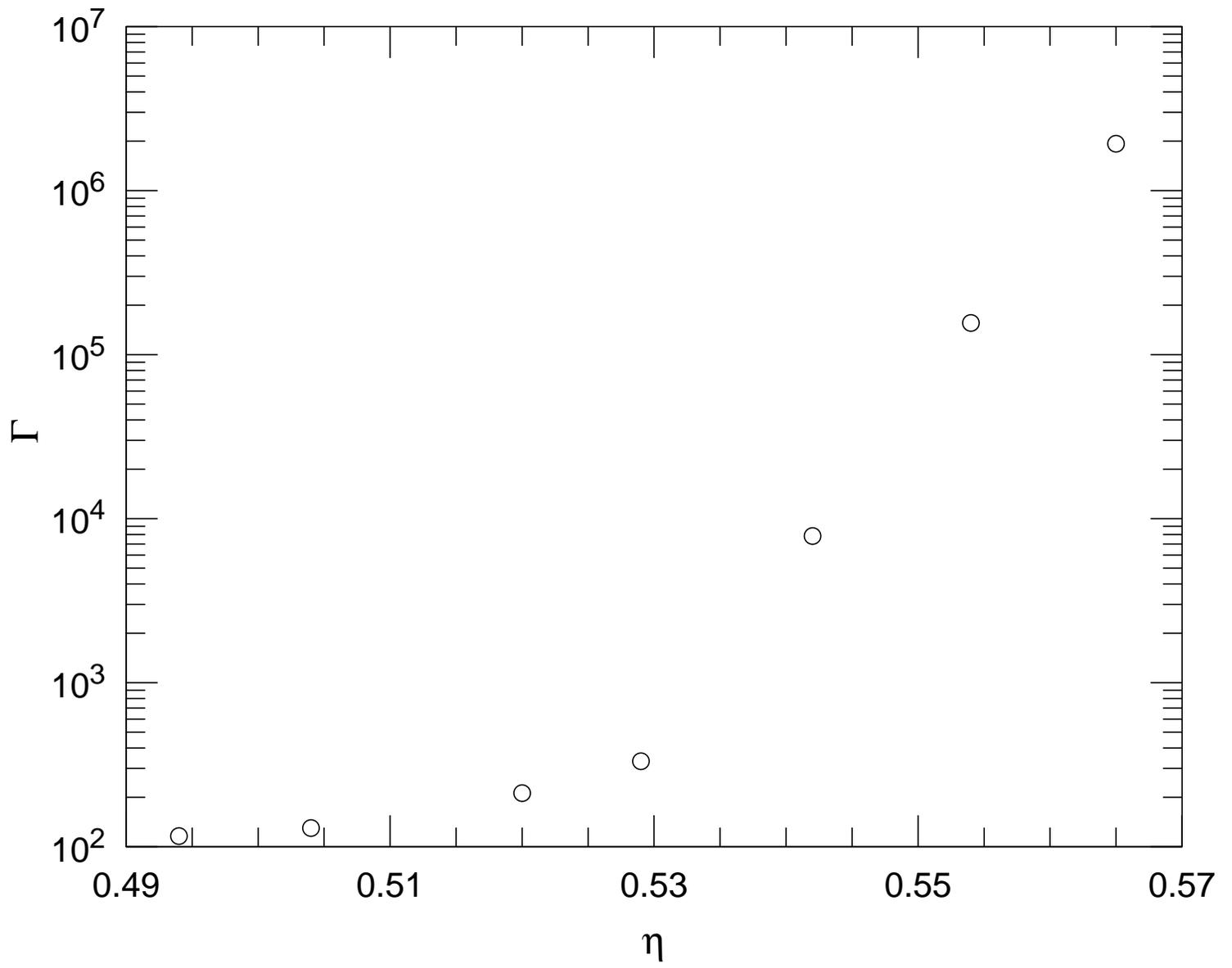
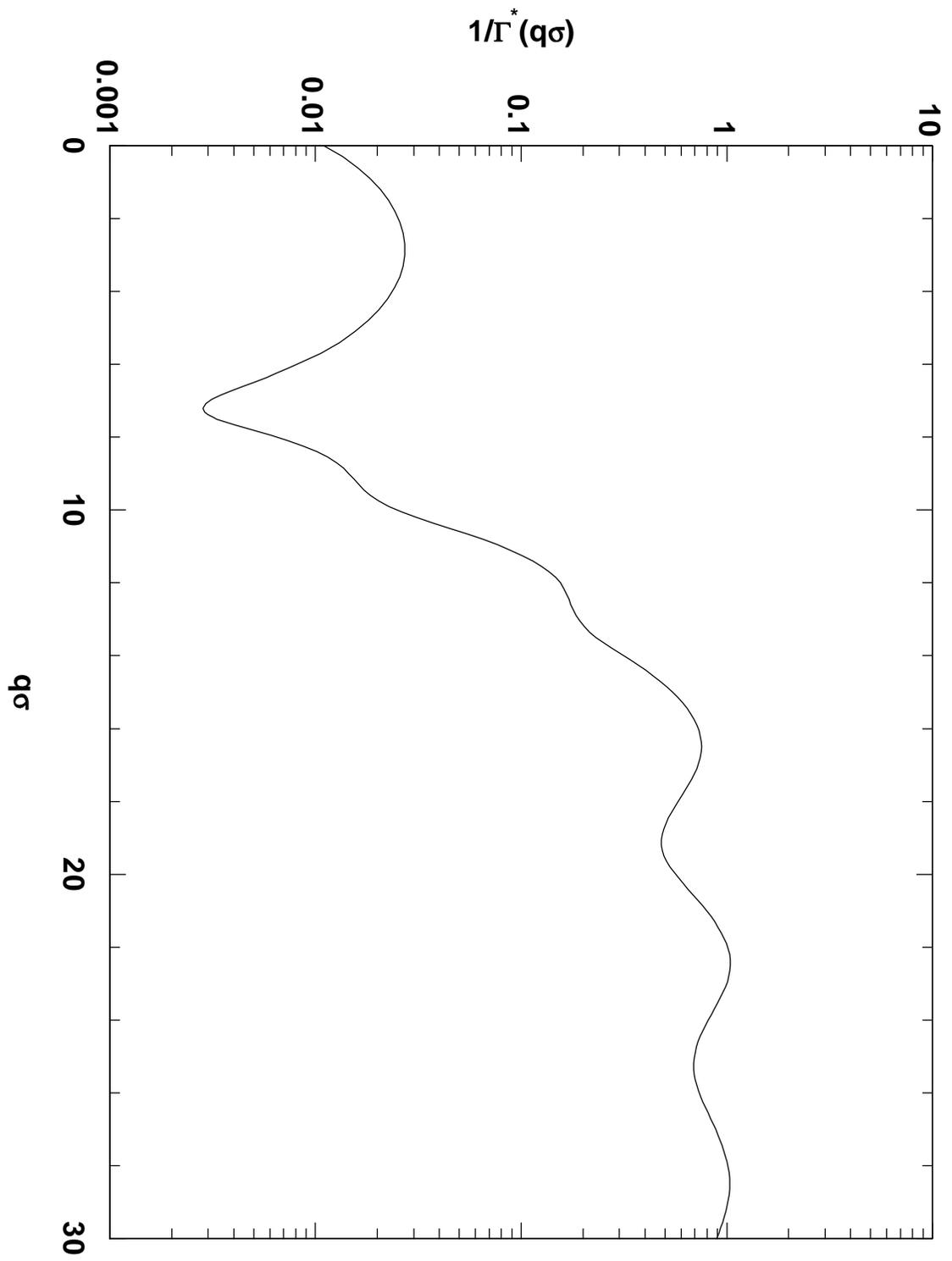
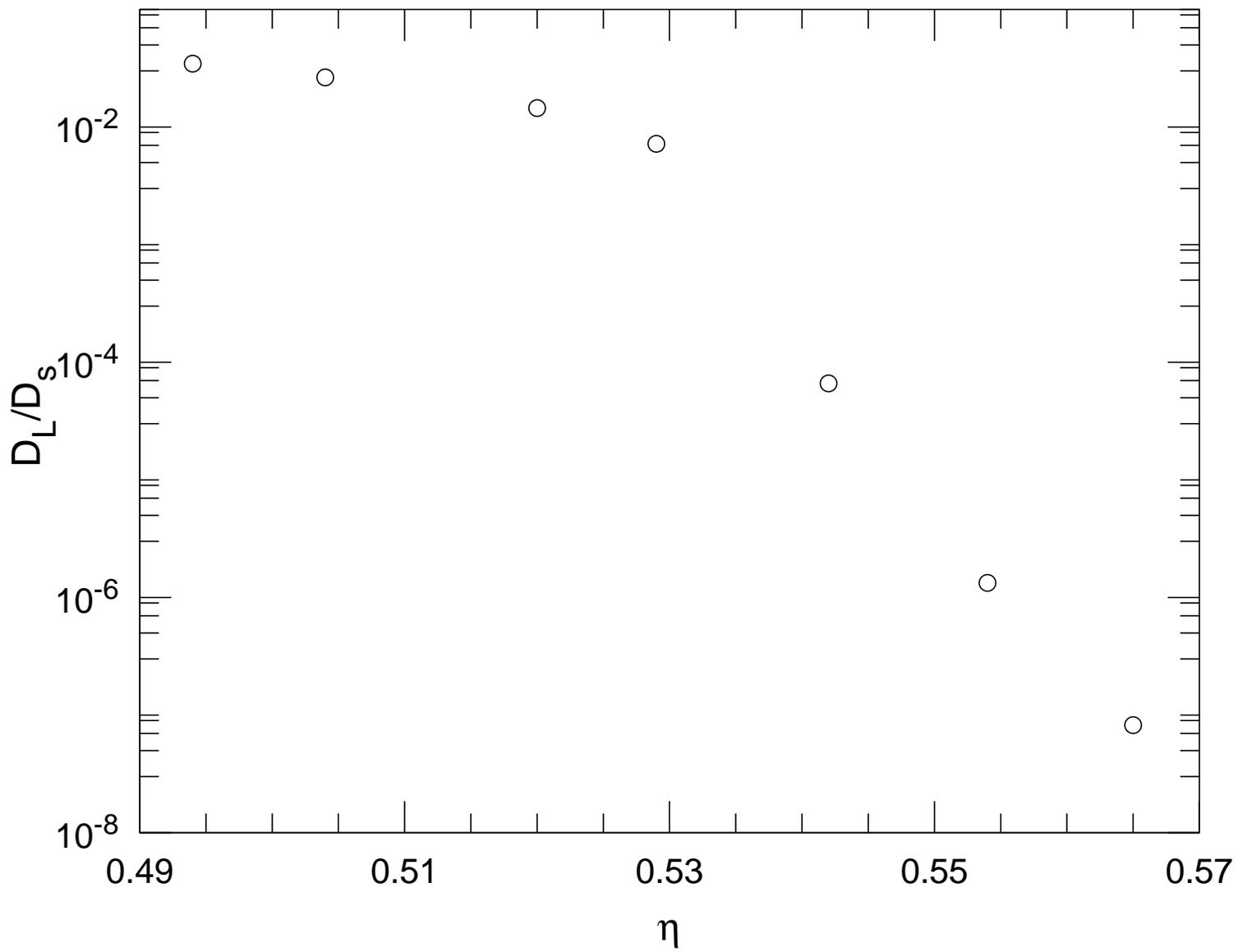
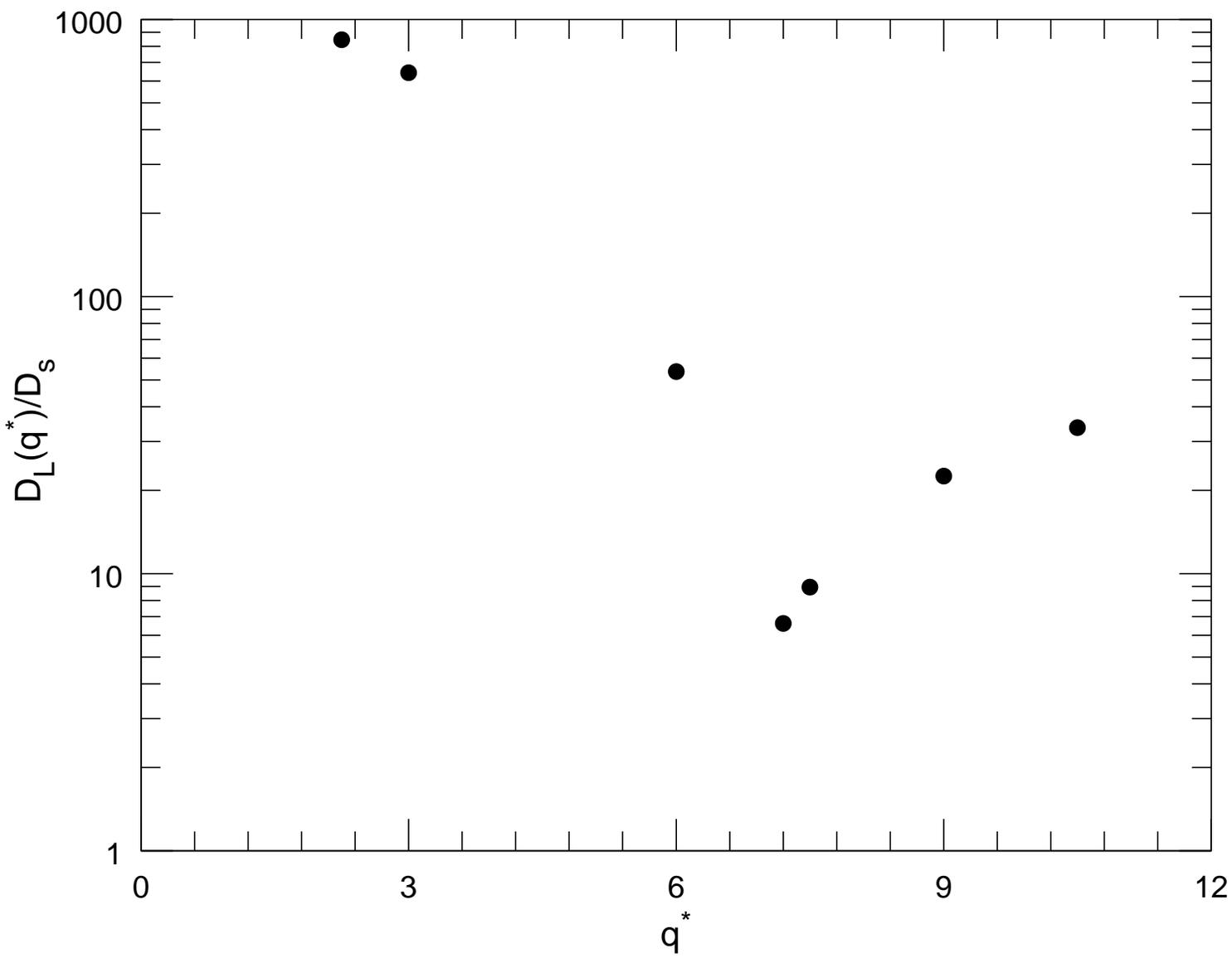
\end{document}